\documentclass[aps,pra,twocolumn,nofootinbib,superscriptaddress]{revtex4-1} 
\usepackage{graphicx}% Include figure files
\usepackage{dcolumn} % Align table columns on decimal point
\usepackage{bm}      % bold math
\usepackage{mathtools}
\usepackage{color}
\usepackage{float}
\usepackage{bbm}
\usepackage{amssymb,amsmath,amsthm,amsfonts}
\usepackage{natbib}
\usepackage{circuitikz}
\usepackage{tikz}
\usetikzlibrary{patterns}
\usepackage{hyperref}

\newcommand{\ket}[1]{\left| #1 \right\rangle}

\begin{document}

\footnotetext{This manuscript has been authored by UT-Battelle, LLC, under Contract No. DE-AC0500OR22725 with the U.S. Department of Energy. The United States Government retains and the publisher, by accepting the article for publication, acknowledges that the United States Government retains a non-exclusive, paid-up, irrevocable, world-wide license to publish or reproduce the published form of this manuscript, or allow others to do so, for the United States Government purposes. The Department of Energy will provide public access to these results of federally sponsored research in accordance with the DOE Public Access Plan.}

\title{The Hybrid Topological Longitudinal Transmon Qubit}
\author{Alec Dinerstein}
\affiliation{Computational Sciences and Engineering Division, Oak Ridge National Laboratory, Oak Ridge, TN 37831}
\author{Caroline S. Gorham}
\affiliation{Computational Sciences and Engineering Division, Oak Ridge National Laboratory, Oak Ridge, TN 37831}
\author{Eugene F. Dumitrescu}
\email{dumitrescuef@ornl.gov}
\affiliation{Computational Sciences and Engineering Division, Oak Ridge National Laboratory, Oak Ridge, TN 37831}

\begin{abstract}

We introduce a new hybrid qubit consisting of a Majorana qubit interacting with a transmon longitudinally coupled to a resonator. To do so, we equip the longitudinal transmon qubit with topological quasiparticles, supported by an array of heterostructure nanowires, and derive charge- and phase-based interactions between the Majorana qubit and the resonator and transmon degrees of freedom. Inspecting the charge coupling, we demonstrate that the Majorana self-charging can be eliminated by a judicious choice of charge offset, thereby maintaining the Majorana degeneracy regardless of the quasiparticles spatial arrangement and parity configuration. We perform analytic and numerical calculations to derive the effective qubit-qubit interaction elements and discuss their potential utility for state readout and quantum error correction. Further, we find that select interactions depend strongly on the overall superconducting parity, which may provide a direct mechanism to characterize deleterious quasiparticle poisoning processes.

\end{abstract}
\maketitle

\section{Introduction}
According to state of the art resource estimates, accurate fault-tolerant quantum computations, of even the simplest non-trivial problems, will require millions of qubits and days of compute time\cite{Kivlichan2020}. Intrinsically topological qubits are promising candidates for reducing the burdensome resource requirements needed for quantum error correction to deliver fault-tolerant quantum computation\cite{Litinski2017,Karzig2017}. Coupling topological with unprotected superconducting qubits is therefore an attractive methodology for scaling up error correcting systems\cite{Xue2015}.  

A variety of superconducting qubits exist and for some, including Cooper pair box and flux qubits, a theory for coupling with topological Majorana modes has been explored\cite{Hassler2010,Hassler2011,Jiang2011,Bonderson2011,Zhang2013}. Recently, a new scalable architecture has been developed that is based on a {\em longitudinally interacting} transmon qubit, which supports a universal gate set \cite{Billangeon2015, Royer2017, Richer2016, Richer2017} while simultaneously implementing a higher fidelity, with respect to dispersive interactions, non-demolition readout\cite{Didier2015}. Improvements in readout fidelity, due to the longitudinal interaction, are important since assignment errors in measurement readout are the dominant error sources, with errors being three to four orders of magnitude larger than that of single qubit gates\cite{Bravyi2020}, which hamper fault tolerance.

We therefore address the open question of combining these two important qubit platforms by developing a theory for the hybrid topological-longitudinal transmon (TLT) qubit. To do so we derive topological-transmon and topological-resonator couplings and discuss the use of the resulting interactions for quantum gates and readout. We also find that various coupling configurations strongly depend on the parity of the Majorana qubit, which could be used to experimentally probe quasiparticle poisoning properties and to control the Majorana qubits parity sector. 

Our work is organized as follows. We first re-introduce a transmon qubit longitudinally coupled to a resonator, which can be realized with a simple inductively shunted circuit design. We then introduce Majorana excitations, supported by a network of one-dimensional (1D) heterostructures, and derive their couplings to both the resonator and transmon. In order to determine the relative strengths of analytically found couplings, we compare the relevant energy scales and compare with numerical results obtained from projecting into the low-energy (qubit) manifold of the transmon. In examining the couplings, we report on the possibility of generating a strong dependence of the interaction terms on the topological qubits' fermion parity. We then conclude by discussing potential applications and open questions.  

\section{Inductively shunted longitudinally coupled transmon}

\begin{figure}[hbt!]
    \centering
    \begin{circuitikz}[x=1cm,y=1cm]
    \filldraw[black] (2,1) circle (2pt) node[anchor=north](1){a};
    \draw (2,1) -- (2,2) to[barrier] (2,3);
    \filldraw[black] (2,3.25) circle (0pt) node[anchor=south, label={[label distance=-0.4cm,text depth=2ex,rotate=90]left:$kE_J$}](2){\vdots};
    \draw (2,4) to[barrier] (2,5) -- (2,6) -- (4,6);
    \filldraw[black] (4,6) circle (2pt) node[anchor=south](3){c};
    \draw (4,6) -- (6,6) -- (6,5) to[barrier] (6,4);
    \filldraw[black] (6,3.25) circle (0pt) node[anchor=south, label={[label distance=-0.4cm,text depth=2ex,rotate=90]left:$kE_J$}](4){\vdots};
    \draw (6,3) to[barrier](6,2) -- (6,1);
    \filldraw[black] (6,1) circle (2pt) node[anchor=north](5){b};
    \draw (6,1) to[barrier, name=eq] (2,1);
    \node[above, xshift=2pt, yshift=14pt] at (eq.n) {$E_{J_q}$};
    \draw (1,2) -- (3,2) to[L,l_=$L$] (3,5) -- (1,5) to[C, l_=$C$] (1,2);
    \draw (5,2) -- (7,2) to[L, l_=$L$] (7,5) -- (5,5) to[C, l_=$C$] (5,2);
    \draw (3,1) -- (3,0) to[C, l_=$C_q$] (5,0) -- (5,1);
    
    \draw (2.4,3.5) circle (.2 cm);
    \filldraw[black] (2.4,3.5) circle (2pt) node[anchor=south, label={[label distance=-0.5cm,text depth=5ex,rotate=0]left:$\Phi_x$}](5){};
    \draw (6.4,3.5) circle (.2 cm);
    \filldraw[black] (6.4,3.5) circle (2pt) node[anchor=south, label={[label distance=-0.5cm,text depth=5ex,rotate=0]left:$\Phi_x$}](6){};
\end{circuitikz}
    \caption{A set of k Josephson junctions, shunted by capacitors and inductors (left and right branches) connected in series with a capacitively shunted Josephson junction (bottom branch) realizes transmon qubit, along the bottom branch, which is longitudinally coupled to a harmonic resonator as described by Eqs.~\ref{eq:L_abc},\ref{eq:H_ab}}
    \label{fig:circuit1}
\end{figure}
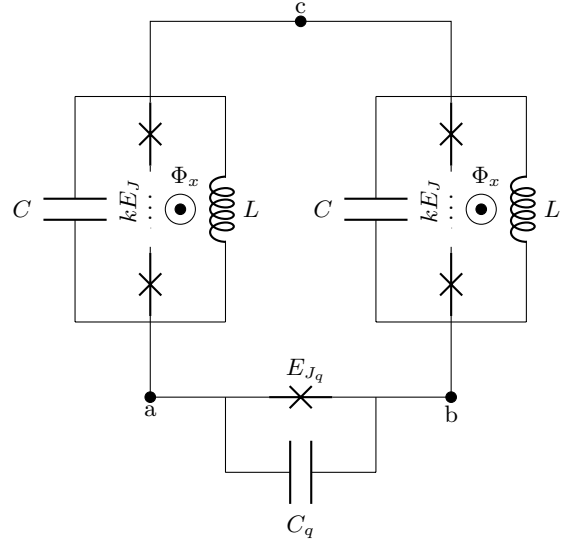

Our construction begins with the longitudinally coupled transmon which was previously introduced in Ref.~\cite{Richer2016} and is illustrated in Fig.~\ref{fig:circuit1}. The systems Lagrangian is $\mathcal{L} = \mathcal{T}-\mathcal{V}$ where
\begin{eqnarray}
\label{eq:L_abc}
\mathcal{T}&=& \left(\frac{\Phi_0}{2\pi}\right)^2 [ \frac{C}{2} ( (\dot{\varphi}_a-\dot{\varphi}_c)^2+(\dot{\varphi}_b-\dot{\varphi}_c)^2 ) \nonumber \\ 
&+&  \frac{C_q}{2} (\dot{\varphi}_a-\dot{\varphi}_b)^2 ] \nonumber \\
\mathcal{V} &=& \frac{1}{2L}\left( (\varphi_a- \varphi_c)^2 +(\varphi_b- \varphi_c)^2 \right) - E_{J_q} \cos{(\varphi_a-\varphi_b)}\nonumber \\ &-& k E_J (\sin{\frac{\varphi_a-\varphi_c}{k}} + \sin{\frac{\varphi_b-\varphi_c}{k}}).
\end{eqnarray}

Here $\varphi_i$ denotes the superconducting phase on the $i$th island, $L$ is an inductance, $C_q$ and $C$ are capacitances, and $E_J, E_{J_q}$ denote Josephson junction energies, related to the junction critical current $I_c$ by $E_J = \Phi_0 I_c/2\pi$.  As shown in Fig.~\ref{fig:circuit1}, in order to further linearize the resulting interactions, the inductively shunted Josephson elements may be further broken into $k$ identical Josephson junctions, i.e. with a constant phase drop across each junction. Furthermore, due to the threading of a flux $\Phi_x=k \Phi_0/4$ where $\Phi_0$ is the flux quanta, the inductively shunted Josephson couplings become {\em sinusoidal} functions of the phase difference. Without loss of generality we take $k=1$.

This system's description is simplified by considering flux quantization and an appropriate change of basis. First, we transform into the basis described by the generalized coordinates which are i) the relative phase difference, $\varphi_q = \varphi_a-\varphi_b$, and ii) the center of mass, $\varphi_+ = \varphi_a+\varphi_b$. In the absence of $E_J$, the $\varphi_q$ coordinate is decoupled from the system and  behaves as an anharmonic resonator, i.e. a transmon, which is discussed in detail below. Transforming again, the coordinate $\varphi_r= \varphi_+ - 2 \varphi_c$ describes a harmonic resonator. The remaining trivial total center of mass coordinate $\varphi_a+\varphi_b+\varphi_c$ is irrelevant as it is eliminated through the flux quantization condition $\sum_{i\in l} \Phi_i - \Phi_{ext} = N \Phi_0$ where $l$ denotes the entire loop. 

% make sure that canonical conjugate variables are correct
Defining $N_i = \frac{1}{\hbar}\frac{\partial \mathcal{L}}{\partial \dot{\varphi_i}}$ as the variable canonically conjugate to $\varphi_i$, and Legendre transforming, we write the Hamiltonian $\mathcal{H} = \sum_{i} N_i \dot{\varphi}_i - \mathcal{L}$
\begin{eqnarray}
\label{eq:H_ab}
\mathcal{H} &=& \frac{\left(2e N_r\right)^2}{C} + \frac{\left(2e N_q\right)^2}{C_\Sigma} + \left(\frac{\Phi_0}{2\pi}\right)^2 \frac{1}{4L}\left(\varphi_q^2 + \varphi_r^2 \right) \nonumber \\
  &-& E_{J_q} \cos{(\varphi_q)} - 2 kE_J (\sin{(\varphi_q/2k)}\cos{(\varphi_r/2k)})
\end{eqnarray}
where $C_\Sigma = 2C_q + C$ and, in the last equation, we have defined the total (and differential) charge operators $N_{r(q)} = N_a \pm N_b$ as above. 

\subsection{Transmon Qubit}
Defining the charging energy $E_{C_q} = \frac{e^2}{C_\Sigma}$ and the tunneling energy $E_{J_q}^* = E_{J_q} + \left( \frac{\Phi_0}{2\pi} \right)^2 \frac{1}{2L}$, the inductively shunted transmon Hamiltonian becomes
\begin{eqnarray}
\label{eq:H_qqqq}
\mathcal{H}_q = 4 E_{C_q} N_q^2 - \frac{ {E_{J_q}^*}^2 }{2}\varphi_q^2 +\frac{{E_{J_q}^*}^4 }{24}\varphi_q^4 + \mathcal{O}(\varphi_q^6). 
\end{eqnarray}
We may go into a second quantization picture by expressing our canonically conjugate variables in terms of field operators $c,c^\dagger$ as $N_q = \frac{1}{2}\sqrt[4]{\frac{E_{J_q}^*}{2E_{C_q}}}i \left( c^\dagger - c \right)$ and $\varphi_q = \sqrt[4]{\frac{2E_{C_q}}{E_{J_q}^*}}\left( c^\dagger + c \right)$. Expanding up to fourth order in $\varphi$, and performing first order perturbation theory results in a duffing oscillator Hamiltonian
\begin{equation}
    \mathcal{H}_q = \hbar \omega_q c^\dagger c  + \delta c^\dagger c (c^\dagger c+1) 
\end{equation}
where $\omega_q = \frac{\sqrt{8E_{C_q}E^*_{J_q}}}{\hbar}$ is the transmon frequency and $\delta=-E_{C_q}/2$ is the absolute anharmonicity. With this anharmonicity one may coherently and individually address the lowest two states, which form the basis for a qubit. 

In order to accommodate a variety of Majorana configurations, each with distinct couplings, we shall later take $E_{J_q}$ to be a \emph{split} junction. This also allows some additional control over the transmon energetics via by modulating the Josephson energy as $E_{J_q} \rightarrow E_{J_q} \cos{\pi \Phi_q/\Phi_0}$\cite{Makhlin2001}, which depends on an external flux $\Phi_q$ as illustrated Fig.~\ref{fig:coupling}.

\subsection{Resonator}
For the resonator, we again re-express the energy contributions from the inductance and capacitance of the circuit.  The charging energy across the two equivalent capacitors in parallel is $E_{C_r} = e^2/C$ and the inductive energy is $E_{L_r} = \left( \frac{\Phi_0}{2\pi} \right)^2 \frac{1}{4L}$. Again, we consider the second quantized form through a standard change of basis to field operators $a,a^\dagger$ through $N_r = \frac{1}{2}\sqrt[4]{\frac{E_{L_r}}{2E_{C_r}}}i \left( a^\dagger - a \right)$ and $\varphi_r = \sqrt[4]{\frac{2E_{C_r}}{E_{L_r}}}\left( a^\dagger + a \right)$. The resulting Hamiltonian, ignoring the zero-point energy, is that of a harmonic oscillator $\mathcal{H}_r = \hbar \omega_r a^\dagger a$  with a transition frequency $\omega_r = \frac{\sqrt{4E_{C_r}E_{J_r}}}{\hbar}$. 

\subsection{Transmon-Resonator Interaction}
To second order in $\varphi_q$ and $\varphi_r$ , the qubit-resonator interaction term is $-\frac{E_J}{8k^2}\varphi_q^2\varphi_r$ which renders our system longitudinally coupled. That is, the transmon operator appearing in the interaction is diagonal. Note that the resonator still interacts in the conventional transverse sense. Putting this into the second quantization representation, and taking first order perturbation theory, we have 
\begin{equation}
\mathcal{H}_{qr} = \hbar g (c^\dagger c + \frac{1}{2})(a^\dagger + a)
\end{equation}
where $g = -\frac{E_J}{4k^2\sqrt{\hbar}} \sqrt{ \frac{2E_{C_q}}{E_{J_q}^*}} \frac{\pi}{\Phi_0} \sqrt[4]{\frac{L}{C}}$  is the longitudinal coupling constant. 

\section{Majorana Coupling}

We now introduce and couple topological Majorana quasiparticles into the superconducting circuit. The Majorana quasiparticles obey anti-commutation relations defined by the Clifford algebra $\{\gamma_i,\gamma_j\}=2\delta_{ij}$ and, given a pair of Majoranas $\gamma_i,\gamma_j$, one may define a conventional Dirac fermion through $c^\dagger = (\gamma_i -i \gamma_j)/2, \; c = (\gamma_i + i \gamma_j)/2$. The Hilbert space of this conventional fermion is described by the usual fermionic number operator $c^\dagger c = (1+i\gamma_i\gamma_j)/2$ with eigenvalues $0,1$ corresponding to the empty and occupied fermionic Fock states.

Establishing the minimal topological qubit involves two pairs of Majoranas which define a four-dimensional fermionic manifold. The $\mathbb{Z}_2$ superconducting fermionic parity constraint, with the $\pm$ degree of freedom referring to the even and odd parity sectors, reduces the size of the fermionic manifold by a factor of two. The resulting two-dimensional space, either $\{ \ket{00}, \ket{11} \}$ or $\{ \ket{01}, \ket{10} \}$, serves as a qubit. The four Majoranas, $\{ \gamma_1, \gamma_2, \gamma_3, \gamma_4 \}$, realize an $\frak{su}(2)$ algebra and we choose the convention $i\gamma_1\gamma_2 \rightarrow Z, i\gamma_1\gamma_3 \rightarrow Y, i\gamma_2\gamma_3 \rightarrow X$.

\begin{figure}[t!]
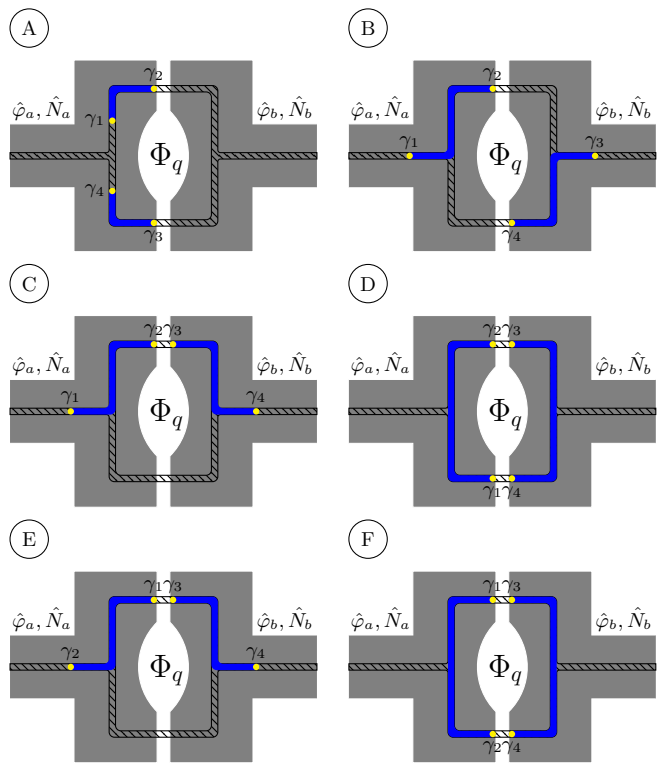

    \centering
    \include{SplitJJ}
    \caption{Majorana configurations in a zoomed in view of the $J_q$ (split) Josephson junction separating nodes a and b. Panels (A) and (B) illustrate configurations coupled by charging effects. The resonator's response depends on the location, state, and parity of the nanowires. In addition, panels (C), (D), (E), and (F) are coupled through the fractional Josephson effect and $\alpha_{ij}$ denotes a spatially dependent overlap integral controlling the interaction strength.}
    \label{fig:coupling}
\end{figure}

Topological Majorana quasiparticles could, for example, be realized by fabricating a variety of nanowire heterostructures\cite{Lutchyn2010, Lutchyn2010, Alicea2011, Mourik2012, Vaitiekenas2020} on top of the underlying superconducting circuit which realizes the transmon and resonators of Fig.~\ref{fig:circuit1}. As per Ref.~\cite{Alicea2011}, we consider a Majorana racetrack architecture in which Majorana can be dragged along a predetermined path indicated by the hatched tubes in Figure~\ref{fig:coupling}. Doing so induces a direct coupling between the topological quasiparticles, in terms of charge\cite{Hassler2011} and phase degrees of freedom.  In such a setup, the positions of the Majoranas can be controlled, for example by electrostatic top-gates depleting electrons from certain regions of the nanowires. As illustrated by Fig.~\ref{fig:coupling}, we utilize a tuning fork shaped nanowire racetrack, a simple extension of the T-junction, surrounding each side of the $J_q$ split junction. This allows one to manipulate the positions of, braid, and fuse the Majoranas. As we discuss below, the control afforded by this design allows us to (de-)couple the Majorana qubit in a variety ways.

Due to the exponential decay of the Majorana interactions, $i \gamma_j \gamma_k \propto e^{-l/\xi}$ where $l$ is the distance between two Majoranas $(j \text{ and } k)$ and $\xi$ is the superconducting coherence length, we ignore coupling by direct Majorana overlap except when explicitly introduced in Section~\ref{subsec:phase}.

\subsection{Charge Coupling}
\begin{table*}[hbt!]
    \begin{center}
        \begin{tabular}{||c | c | c | c | c||} 
            \hline
            & Parity & Coupling Strength & Analytical Value ($h$ GHz) & Numerical Value ( $h$ GHz)\\  [0.5ex] 
            \hline\hline
             $\mathcal{H}_{\gamma}$   & +1 & $4(E_{C_r} + E_{C_q})\sigma_z^{(\gamma)}$ & 1.969 & 1.969 \\
            \hline
            $\mathcal{H}_{r\gamma}$   & +1 &  $\sqrt[4]{E_{L_r}E_{C_r}^3/2} \sigma_y^{(r)}\sigma_z^{(\gamma)}$  & 0.7726  & 0.7726 \\ 
            \hline
            $\mathcal{H}_{q\gamma}$ & +1 & $\sqrt[4]{E_{J_q}^*E_{C_q}^3/2} \sigma_y^{(q)}\sigma_z^{(\gamma)}$ & 0.4727  & 0.4705  \\ [1ex] 
            \hline
        \end{tabular}
        \caption{Charge coupling interactions for configuration (A) in Fig.~\ref{fig:coupling}. The parameter values we used in our calculation are $L$ =  4.5 nH, $C$ = 114 fF, $C_q$ = 70 fF, $E_{J}$ = $h$ 10.00 GHz, $E_{J_q}$ = $h$ 10.00 GHz, which correspond to the values given in Ref.~\citenum{Richer2017}. The resulting charging, inductive, and Josephson energy scales are therefore $E_{C_r}$ = $h$ 0.1525 GHz, $E_{C_q}$ = $h$ 0.3398 GHz, $E_{L_r}$ = $h$ 18.17 GHz, and $E_{J_q}^*$ = $h$ 28.17 GHz.}
        \label{table:1}
    \end{center}
\end{table*}
\begin{table*}[hbt!]
    \begin{center}
        \begin{tabular}{||c | c | c | c | c||} 
            \hline
            & Parity & Coupling Strength & Analytical Value ($h$ GHz) & Numerical Value ($h$ GHz)\\  [0.5ex] 
            \hline\hline
             $\mathcal{H}_{\gamma}$   & +1 & $4E_{C_r}\sigma_z^{(\gamma)}$ & 1.359 & 1.359 \\
            \hline
            $\mathcal{H}_{r\gamma}$   & +1 &  $\sqrt[4]{E_{L_r}E_{C_r}^3/2} \sigma_y^{(r)}\sigma_z^{(\gamma)}$ & 0.7726  & 0.7726 \\ 
            \hline
            $\mathcal{H}_{q\gamma}$ & -1 & $\sqrt[4]{E_{J_q}^*E_{C_q}^3/2} \sigma_y^{(q)}\sigma_z^{(\gamma)}$ & 0.4727  & 0.4705  \\ [1ex] 
            \hline
        \end{tabular}
        \caption{Charge coupling interactions for the configuration illustrated in Fig.~\ref{fig:coupling} (B). System parameters are the same as those reported in Table ~\ref{table:1}.}
        \label{table:2}
    \end{center}
\end{table*}

Despite a topological fermionic orbital being encoded non-locally when the Majoranas are spatially well separated, the presence or absence of an additional charge contributes to the superconducting islands charging energy. We denote the Majorana-encoded fermion occupation numbers on each superconducting island as $m_i = \sum_{j,k \in i} (1+i\gamma_j\gamma_k)/2$ where $j,k \in i$ signifies that $\gamma_j,\gamma_k$ reside on island $i$ (with islands $a,b$ illustrated in Fig.~\ref{fig:circuit1}). With this notion, we may define the differential and total charge operators as $m_q=(m_a-m_b)/2$ and $m_r = (m_a+m_b)/2$. Note that a factor of two appears in the denominator, which accounts for the fact that we have expressed our charging energy in terms of the fundamental Cooper pair charge $e^*=2e$. The Majorana occupation numbers then couple to the differential and resonator charges as
\begin{eqnarray}
\label{eq:charge_coupling}
N_q & \rightarrow & N_q + m_q \\  \nonumber
N_r & \rightarrow & N_r + m_r.
\end{eqnarray}

% Majorana charge offset
Inserting these charge transformations into Eq.~\ref{eq:H_ab}, one can see that not only will the Majoranas couple to the transmon but they will also self-interact through a capacitive coupling. In order for the topological qubit to remain insensitive to external perturbations the Majorana manifold should be kept degenerate. As previously shown by Ref.~\cite{Fu2010}, the Majorana degeneracy can be maintained by a judicious charge offset $n_g$ which enters the charging Hamiltonian as $\mathcal{H}_C = (2\hat{N} - n_g)^2/(2 C)$. Ref.~\citenum{Fu2010} considers a single superconducting island hosting a pair of Majorana modes which produce a non-local fermionic mode $\{\ket{0}, \ket{1} \}$. Setting the background offset charge to $n_g=(2N+1/2)$ creates a degeneracy between the $\ket{N-1}\ket{1}$ and $\ket{N}\ket{0}$ states where we have taken a tensor product over the Cooper pair space and that of the non-local fermionic mode.

Likewise, we now show that for all of our configurations an appropriate charge offset may preserve the Majorana degeneracy. Subtracting an offset charge $n_g^{(r,q)} = n_g^{(a)} \pm n_g^{(b)}$ from Eq.~\ref{eq:charge_coupling}, and expanding the capacitive terms of Eq.~\ref{eq:H_ab}, one can write the Majorana contribution to the energy as,
\begin{equation}
\label{eq:majorana_charging}
    H_\gamma = 4\left[ E_{C_r}(m_r-n_g^{(r)})^2 + E_{C_q}(m_q-n_g^{(q)})^2 \right].
\end{equation}

Consider the first configuration illustrated in Fig.~\ref{fig:coupling} panel (A), where both pairs of Majoranas are located on island $a$ such that $m_q=m_r=m_a$. In the odd parity sector, for which $\gamma_3\gamma_4 = - \gamma_1\gamma_2$, the Majorana operators destructively interfere and $m_a$ reduces to a constant, thus preserving the Majorana degeneracy. In the even parity sector, where $\gamma_3\gamma_4 = \gamma_1\gamma_2$, we get $m_a = (1+i\gamma_1\gamma_2)/2$ such that $(m_q-n_g^{(q)})^2 = (i\gamma_1\gamma_2)^2/4=\pm1/4$ is a Majorana-independent constant given a charge offset $n_g^{(r)} = n_g^{(q)} =n_g^{(a)} = 1/2$. 
 
A similar analysis can be performed for the configuration illustrated in Fig.~\ref{fig:coupling} (B), where $m_r$ and $m_q$ differ depending on the parity sector. For odd parity, $m_r$ remains a constant while $m_q = (i\gamma_1\gamma_2)/2$. This Majorana interaction, going as $m_q^2$, turns into a constant even without a charge offset $n_g^{(q)}=0$. Finally, in the even parity sector, the differential Majorana occupation $m_q$ vanishes while $m_r = (1+i\gamma_1\gamma_2)/2$ and the charge offset of $n_g^{(r)} = n_g^{(q)} =n_g^{(a)} = 1/2$ maintains the Majorana degeneracy as in case (A). Thus, by maintaining $n_g^{(a)} = 1/2$ for even parity and $n_g^{(a)} = 0$ for odd parity, we see that the Majoranas are free to move from one island to another while maintaining a degenerate Majorana subspace. With these charging conditions, we may now braid the Majoranas through the use of the T-junctions and bring together arbitrary pairs of Majoranas. This will prove useful in implementing the phase couplings discussed in Section~\ref{subsec:phase}.

The remaining interactions couple the Majorana qubit to the resonator and transmon, and are of the form $ \mathcal{H}_{r\gamma} = 8 E_{C_r} N_r m_r $ and $ \mathcal{H}_{q\gamma} = 4 E_{C_q} N_q m_q $. From the Majorana charging energy discussion, and field operator definitions of $N_i$, we see that the charging interactions are of the form ${\sigma_y^{(r)}}\sigma_z^{(\gamma)}$ and ${\sigma_y^{(q)}}\sigma_z^{(\gamma)}$ -- where the Pauli superscript denotes the interacting subsystem. Table~\ref{table:1} presents the analytical expression for this interaction strength, where we have projected into the lowest two energy levels for all degrees of freedom. The analytical value corresponds to parameters arising from a low order expansion of the Josephson energy with respect to the original Fock states. We have also calculated the exact results arising from: first diagonalizing the transmon Hamiltonian using a Fock state cutoff of $N=5$ and then computing the matrix elements that connect the $\pm1$ eigenstates of the $\sigma^{(q)}_x, \sigma^{(q)}_y, \sigma^{(q)}_z$ operators defined using the numerically determined eigenvectors\cite{qutip}. We see that this gives rise to corrections which are about $1\%$ of the total interaction strength. 

Note that these interactions are {\em precisely} of the form required to implement a longitudinal readout of the topological qubit\cite{Didier2015}, i.e. diagonal with respect to the topological qubit and transverse with respect to the resonator. Parametric modulation of the interaction strength, at the frequency of the resonator cavity, is needed to realize longitudinal readout. While parametrically tuning such couplings, e.g. the charging energy, is non-trivial, recent developments in piezo-mechanics for superconducting architectures provides an avenue to realize the required modulation\cite{Han2020}. 

% summarize findings in table
\begin{table*}[t!]
    \begin{center}
        \begin{tabular}{||c | c | c | c | c||} 
            \hline
          Configuration  & Parity &  Coupling Strength & Analytical Value ($h$ GHz) & Numerical Value ($h$ GHz)\\  [0.5ex] 
            \hline\hline
            $(C)$ & +1 & $\Delta \ \sqrt[]{E_{C_q}/32E_{J_q}^*} \sigma_z^{(q)}\sigma_x^{(\gamma)}$ & 0.3145 & 0.3161   \\
            \hline
            $(C)$& -1 & $\Delta \ \sqrt[]{E_{C_q}/32E_{J_q}^*} \sigma_z^{(q)}\sigma_x^{(\gamma)}$ & 0.3145 & 0.3161   \\
            \hline
            $(D)$ & +1 & $\Delta \ \sqrt[]{E_{C_q}/8E_{J_q}^*} \sigma_z^{(q)}\sigma_x^{(\gamma)}$ & 0.6290  & 0.6321    \\
            \hline
            $(E)$ & +1 & $\Delta \ \sqrt[]{E_{C_q}/32E_{J_q}^*} \sigma_z^{(q)}\sigma_y^{(\gamma)}$ & 0.3145 & 0.3161 \\
            \hline
            $(E)$ & -1  & $\Delta \ \sqrt[]{E_{C_q}/32E_{J_q}^*} \sigma_z^{(q)}\sigma_y^{(\gamma)}$ & 0.3145 & 0.3161  \\
            \hline
            $(F)$ & -1 & $\Delta \ \sqrt[]{E_{C_q}/8E_{J_q}^*} \sigma_z^{(q)}\sigma_y^{(\gamma)}$ & 0.6290  & 0.6321 \\ [1ex] 
            \hline
        \end{tabular}
        \caption{Phase coupling interactions for configurations Fig.~\ref{fig:coupling} (C), (D), (E), and (F).}
        \label{table:3}        
    \end{center}
\end{table*}
\subsection{Phase Coupling}
\label{subsec:phase}

In addition to charging effects, Majorana quasiparticles may be coupled in a way which depends on the superconducting phase degrees of freedom. Such a coupling can be achieved by bringing a pair of Majoranas into close proximity across a Josephson junction. If the superconducting phase difference across a Josephson junction is $\varphi_q=\pi$, the superconducting order parameter mimics a mass kink in the Jackiw-Rebbi model and the Majoranas are uncoupled\cite{Kane2013}. On the other hand, if $\varphi_q=0$ the system constitutes a single superconductor and, if fully overlapping, the Majoranas hybridize into a conventional, i.e. gapped, quasiparticle state. This coupling is modeled by the so-called fractional Josephson effect, which is written $H_{FJE} = \Delta \alpha_{23} i \gamma_2 \gamma_3 \cos(\varphi_q/2)$ where $\alpha_{i,j}$ encodes the Majorana's spatial overlap (i.e., $\alpha_{i,j}$ is zero when $\gamma_i$ and $\gamma_j$ are far apart and $\alpha_{i,j}$ is one when they are completely overlapping) and $\Delta$ denotes the topological superconducting gap which we take to be a proximity induced pairing potential directly proportional to the substrate's pairing potential and equal on both Majorana wires. To date, typical topological superconducting nanowire heterostructures have realized proximity-induced superconducting gaps of $\Delta \sim 100 \mu eV$ \cite{Mourik2012,Vaitiekenas2020} which corresponds to $24$ GHz. To determine the effective Majorana-transmon coupling constant, we compute the coupling matrix element by expanding Eq.~\ref{eq:FJE} to second order in $\varphi_q$ and project into the transmon qubit subspace. We then consider each of the Majorana configurations, illustrated in Figure~\ref{fig:coupling}, which are modeled as:
\begin{eqnarray}
\label{eq:FJE}
\mathcal{H}_{q\gamma} =  \left\{
        \begin{array}{l}
           C: \; \Delta(i\gamma_2\gamma_3)\cos{(\varphi_q/2)}\\
           D: \; \Delta(i\gamma_2\gamma_3+i\gamma_1\gamma_4)\cos{(\varphi_q/2)}\\
           E: \; \Delta(i\gamma_1\gamma_3)\cos{(\varphi_q/2)}\\
           F: \; \Delta(i\gamma_1\gamma_3+i\gamma_2\gamma_4)\cos{(\varphi_q/2)}
        \end{array}\right. .
\end{eqnarray}

The resulting interactions, expressed in the Pauli basis, are given in Table~\ref{table:3}. Since the $\varphi_q$ operator is diagonal with respect to the transmon spectrum, or approximately diagonal in the exact case, the transmon interactions go as $\sigma_z^{(q)}$ and the topological component interactions go as $\sigma_x^{(\gamma)}$ or $\sigma_y^{(\gamma)}$, depending on the specific Majoranas which are brought together.

% clifford circuits for QEC
It is important to note that the interactions listed in Table~\ref{table:3} can be utilized to implement Clifford operations on the joint Majorana-transmon subspace which are necessary to implement stabilizer parity check measurements for quantum error correction\cite{Fowler2012}. Specifically, the transmon-Majorana $\sigma_z^{(q)}\sigma_y^{(\gamma)}$ interaction can be used to implement a controlled-NOT gate with additional single-qubit Clifford operations\cite{Maslov2017} which may be realized by local transmon gates and Majorana braiding operations. 

\subsection{Parity Dependence}
Additionally, as highlighted in the second column of Tables ~\ref{table:1}, \ref{table:2}, and \ref{table:3}, the hybrid TLT interactions strongly depend on the overall superconducting parity. Such interactions are of particular interest within the field of topological superconducting physics as they can be used to directly probe the superconducting parity itself. 

For example, consider the situation where the charge offset for configuration Fig.~\ref{fig:coupling} (A) is set to $n_g^{(a)} =1/2$ in order to maintain the Majorana degeneracy. Then, by reducing the charge offset to $n_g^{(a)}=0$ a $\sigma^{(\gamma)}_z$ coupling is introduced for the $+1$ parity sector while the $-1$ parity sector remains uncoupled. Likewise, various phase coupling configurations (see Table~\ref{table:3}) also display a strong parity dependence in the transmon-Majorana interaction. By controlling and monitoring the state of the transmon qubit, e.g. through longitudinal readout in this platform\cite{Didier2015}, the Majorana parity sector can be read out from the parity-dependent transmon dynamics. This provides a technique for directly reading out the total Majorana parity and measuring the deleterious quasiparticle poisoning rates\cite{Rainis2012} which have thus far been detected through parity non-conserving quasiparticle transport\cite{Albrecht2017}. 

\section{Conclusion}

In this work, we have introduced a new TLT hybrid qubit which consists of a Majorana-based topological qubit interacting with an longitudinally-coupled transmon qubit. In doing so, our TLT qubit architecture paves the way for coherently controlling and manipulating quantum information between topological degrees of freedom and the recently developed longitudinally-interacting transmon qubit that is crucially equipped with a non-demolition readout mechanism\cite{Didier2015}. 

Beginning with the circuit construction of Ref.~\cite{Richer2016}, we equip the system with topological quasiparticles supported by a nanowire array lying on top of the superconducting substrate. We have derived charge- and phase-based interactions between the Majorana qubit and the resonator and transmon degrees of freedom. We have demonstrated that the Majorana self-charging can be eliminated by a judicious choice of charge offset, thereby maintaining the Majorana degeneracy regardless of their spatial arrangement and parity configuration. 

We have performed analytic and numerical calculations to derive the effective qubit-qubit interaction elements and have discussed their potential utility in readout and quantum error correction schemes. We have also found that select interactions depend strongly on the overall superconducting parity, which could provide a direct mechanism to probe deleterious quasiparticle poisoning processes and to control the topological parity sector. 

Given the variety of derived interactions and their application, further work is necessary to quantify the quantum information processing protocols that we have discussed for particular circuit realizations. For example, a signal-to-noise analysis can be performed to bound the timescales necessary for readout or state transfer of the topological qubits. Also, a sensitivity analysis for the interactions with respect to fluctuations in system parameters, is needed to determine the expected fidelities for the intra-TLT interactions. 

Before submission we became aware of Ref.~\citenum{Smith2020} which contains some overlap with our current work. 

\section{Acknowledgements}
This research was sponsored by  the  U. S. Department of Energy, Office of Science, Basic Energy Sciences, Materials Sciences and Engineering Division. 

% \appendix
% \section{Majorana Charging Energies}

% \begin{eqnarray}
% \mathcal{H}_{Q}^{(i)} &=& 4E_{C_i} \left(N_i + m_i -n_g^{(i)}\right)^2  \nonumber \\
% &=& \frac{4e^2}{C_i} ((N_i^2 -2N_i n_g^{(i)})  \nonumber \\ 
% &+& (m_i^2 - 2m_i n_g^{(i)}) \nonumber \\ 
% &+& 2 N_i m_i - n_g^{(i)2} ) \nonumber \\ 
% &=& \frac{4e^2}{C_i} ((N_i^2 - n_g^{(i)})^2 + (m_i^2 - n_g^{(i)})^2 \nonumber \\ 
% &+& 2 N_i m_i + n_g^{(i)2} ) \nonumber \\ 
% \end{eqnarray}

\bibliography{main.bib}
\end{document}